# Development of a Real-time Indoor Location System using Bluetooth Low Energy Technology and Deep Learning to Facilitate Clinical Applications

Guanglin Tang, Yulong Yan, Chenyang Shen, Xun Jia, Meyer Zinn, Zipalkumar Trivedi, Alicia Yingling, Kenneth Westover, Steve Jiang

Medical Artificial Intelligence and Automation (MAIA) Laboratory, Department of Radiation Oncology, University of Texas Southwestern Medical Center, Dallas, TX 75235, United States of America

\* Corresponding Author: Guanglin.Tang@utsouthwestern.edu

## Abstract

**Purpose**: An indoor, real-time location system (RTLS) can benefit both hospitals and patients by improving clinical efficiency through data-driven optimization of procedures. Bluetooth-based RTLS systems are cost-effective but lack accuracy because Bluetooth signal is subject to significant fluctuation. We aim to improve the accuracy of RTLS using the deep learning technique.

**Methods**: We installed a Bluetooth sensor network in a 3-floor clinic building to track patients, staff, and devices. The Bluetooth sensors measured the strength of the signal broadcasted from Bluetooth tags, which was fed into a deep neural network to calculate the location of the tags. The proposed deep neural network consists of a Long Short-Term Memory (LSTM) network and a deep classifier for tracking moving objects. Additionally, a spatial-temporal constraint algorithm was implemented to further increase the accuracy and stability of the results. To train the neural network, we divided the building into 115 zones and collected training data in each zone. We further augmented the training data to generate cross-zone trajectories, mimicking the real-world scenarios. We tuned the parameters for the proposed neural network to achieve relatively good accuracy.

**Results**: The proposed deep neural network achieved an overall accuracy of about 97% for tracking objects in each individual zone in the whole 3-floor building, 1.5% higher than the baseline neural network that was proposed in an earlier paper, when using 10 seconds of signals. The accuracy increased with the density of Bluetooth sensors. For tracking moving objects, the proposed neural network achieved stable and accurate results. When latency is less of a concern, we eliminated the effect of latency from the accuracy and gained an accuracy of 100% for our testing trajectories, significantly improved from the baseline method.

**Conclusions**: The proposed deep neural network composed of a LSTM, a deep classifier and a posterior constraint algorithm significantly improved the accuracy and stability of RTLS for tracking moving objects.

# 1. Introduction

Indoor Real-time Location Systems (RTLS) that enable precise position tracking are beneficial in many fields. In industrial manufacturing, an RTLS enables autonomous robots to navigate towards designated locations or other objects whose locations the RTLS tracks simultaneously [1]. In malls and public transit, it allows managers to track assets in the field and analyze and predict customer traffic to provide adequate time to prepare for their arrival [2]. Particularly, in clinical environments, an RTLS can help deliver health care more efficiently, reduce clinical errors, and enhance patient safety by monitoring patients' locations, enabling a rapid response by healthcare providers in the event of a medical emergency [3] [4] [5]. In the Radiation Oncology building (EROC) at the University of Texas Southwestern Medical Center (UTSW), An in-house RTLS is used to improve patient safety by automatically matching the location of patients and their specific devices and accessories before scheduled delivery of radiotherapy to ensure that the right patients are treated and the right devices and accessories are placed. Additionally, historical location records of staff, patients, and equipment can help detect and reduce the effects of radiation exposure and disease infection by identifying who may have been in the exposed area within a given time frame.

Real-time location tracking can be achieved through several methods, including choke-point schemes [6] [7] [8], time-difference–based schemes [9] [10] [11], and signal strength-based schemes [12] [13] [14]. Compared to the other two schemes, the signal strength-based schemes are more cost-effective. One of the techniques that were applied to reduce cost in RTLS is the Bluetooth Low Energy (BLE) [15] [16] [17] [18], which has been widely used in wireless data transmission for years with iterative improvements in protocol and hardware implementation. The battery life of a low-cost (~$10) BLE transmitter can easily span years. The BLE sensors are also inexpensive (~$10 each) and industrially mature with easy connections to computers using common protocols and simple software control. Thus, it is relatively uncomplicated to deploy a reliable RTLS with many BLE devices to enhance signal collection, and thus location accuracy, while minimizing upfront costs and effort spent on software development.

However, most existing BLE-based RTLS solutions suffer from low accuracy, mainly because of the noisy nature of BLE signals due to strong interference [19], which makes the system insufficient for clinical use. Limitation in localization algorithms could also contribute to such low accuracy. For example, because BLE signal strength measured from sensors are not only affected by the distance, but also by the infrastructure and the interference of BLE signals in space [20] [21], the widely used trilateration algorithm, which calculates the location of an object by computing its distances to three reference points, results in biased and unstable location determination.

To attempt to increase the accuracy of BLE-based RTLS, an earlier paper [12] has developed a deep learning algorithm composed of a Convolutional Neural Network (CNN) and an Artificial Neural Network (ANN) to determine the location and compared the results with the traditional methods such as the thresholding method and trilateration method. The paper showed that the proposed deep learning algorithm, so called CNN+ANN, outperformed the traditional methods in terms of accuracy of zone determination, mainly because the deep learning algorithm reduced the infrastructure-induced systematic error, which is unavoidable in traditional methods, by fitting a function to directly map the measured signals (the so-called footprint) to the location and explored measurements from more sensors than traditional methods. They tested the algorithm in a clinic area comprising 21 zones on a single floor and evaluated the accuracy for each zone individually, which, however, did not cover trajectories in which tracked objects move between different zones. When we tried to apply CNN+ANN to track moving objects in the building, we still found that the determined location jumped back-and-force between zones, especially during zone transitions, confusing some of clinical applications. Therefore, additional efforts are needed in developing new algorithms that are more accurate and stable to fulfill the needs of clinical applications.

To address the instability issue, we noticed that the previous method has not fully explored the temporal relation between steps, saying the current location depends on previous locations to certain extent. To account for the temporal relation between steps, we considered using the Long Short-Term Memory (LSTM) [22] architecture, which has been widely used in the fields of natural language processing [23] [24] and video processing [25] [26] because it takes into account memory from historical information when making predictions. Additionally, the role of spatial relation between zones in determining locations has not been fully explored in previous methods either. For example, it is more likely for an object to stay in the current zone or move to a near zone than to a far-away zone from the previous step. Thus, we considered accounting for the known spatial relation between zones using a posterior process to further increase the stability.

In this paper, we developed a deep learning algorithm composed of a LSTM architecture, a deep classifier, and a posterior process, to track moving objects. The algorithm was able to explore the spatial relationship between zones and temporal information of signal measurements, so as to increase the accuracy and stability. We applied our scheme to an RTLS to locate objects in an entire 3-floor clinic building as a more general application scenario.

## 2. Methods

*2.1 System description*

We established a hardware infrastructure as a testbed for RTLS algorithms. In the three-floor EROC building at UTSW, we deployed a fleet of 142 Raspberry Pis, developed by the Raspberry Pi Foundation of United Kingdom, to host 142 USB Bluetooth adapters as Bluetooth sensors. To track a patient or device, we attach to each of them a Beacon Dot tag, which is developed by the Radius Networks, Inc. at Washington DC, United States.

Figure 1 illustrates the workflow and hardware components of the deployed RTLS. A BLE tag rapidly broadcasts signals, of which the strength is measured by the sensor fleet and transferred to a centralized server equipped with the proposed algorithm to use the data to determine the location of the tag. The location results, once available, are broadcasted to downstream applications, such as for clinical workflow optimization and patient safety checkup.

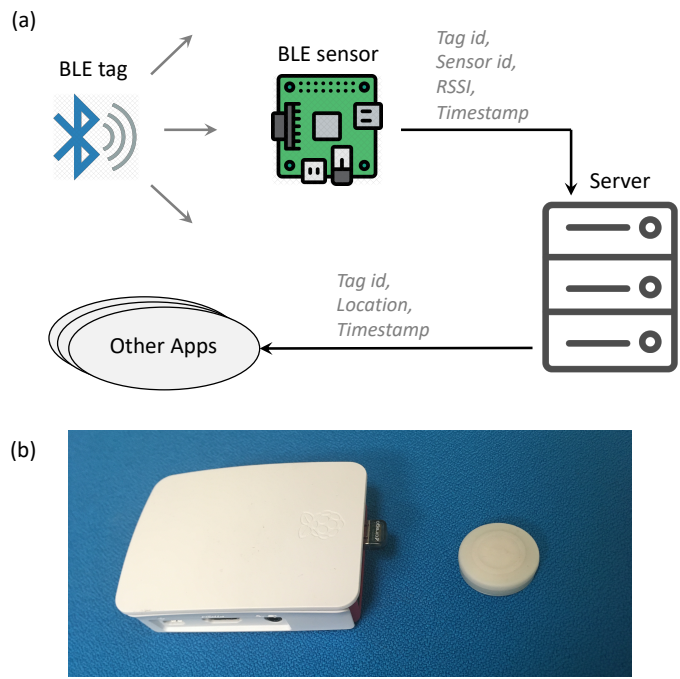

**Figure 1**. (a) Workflow of RTLS deployed in EROC at UTSW. The BLE tag broadcasts BLE signals, of which the strength is measured by the BLE sensors deployed in the building. The latter then transfer the data to the server, on which the proposed algorithm uses the data to determine the location of the tag. The location results are then passed to other apps such as a web application for patients' safety checkup. (b) A Raspberry Pi hosting a USB Bluetooth sensor (left) and a BLE tag beacon (right).

*2.2 Overview of the proposed method*

Based on the installed RTLS hardware infrastructure, we developed a Deep Neural Network composed of a Long Short-Term Memory (LSTM) unit [22] and a deep classifier along with a posterior process to determine the location of a tag with high accuracy and stability. More specifically, we formulated the location determination as a classification problem, in which the building was divided into a number of zones and the measured signals of a tag from the sensor fleet were used to classify its zone using the proposed network. The network was trained with labeled collected data.

In the following sections, we explained in detail the structure, data collection and augmentation, and evaluation of proposed network.

*2.3 Proposed algorithm*

We first divided the building floor map into a number of zones (a total of 115 for EROC) according to their functions, such as exam rooms, treatment rooms, offices, hallways, and waiting areas, which are meaningful for the potential applications of RTLS (see Fig. 2). Then, for a given tag, because of the substantial fluctuation and large portion of missing signals in the acquired signal strength data, we stabilized the signals by averaging the data over a short period of time $\Delta t$. Finally, we use the temporal-averaged signal strengths as input to the proposed network to determine the location of the tag.

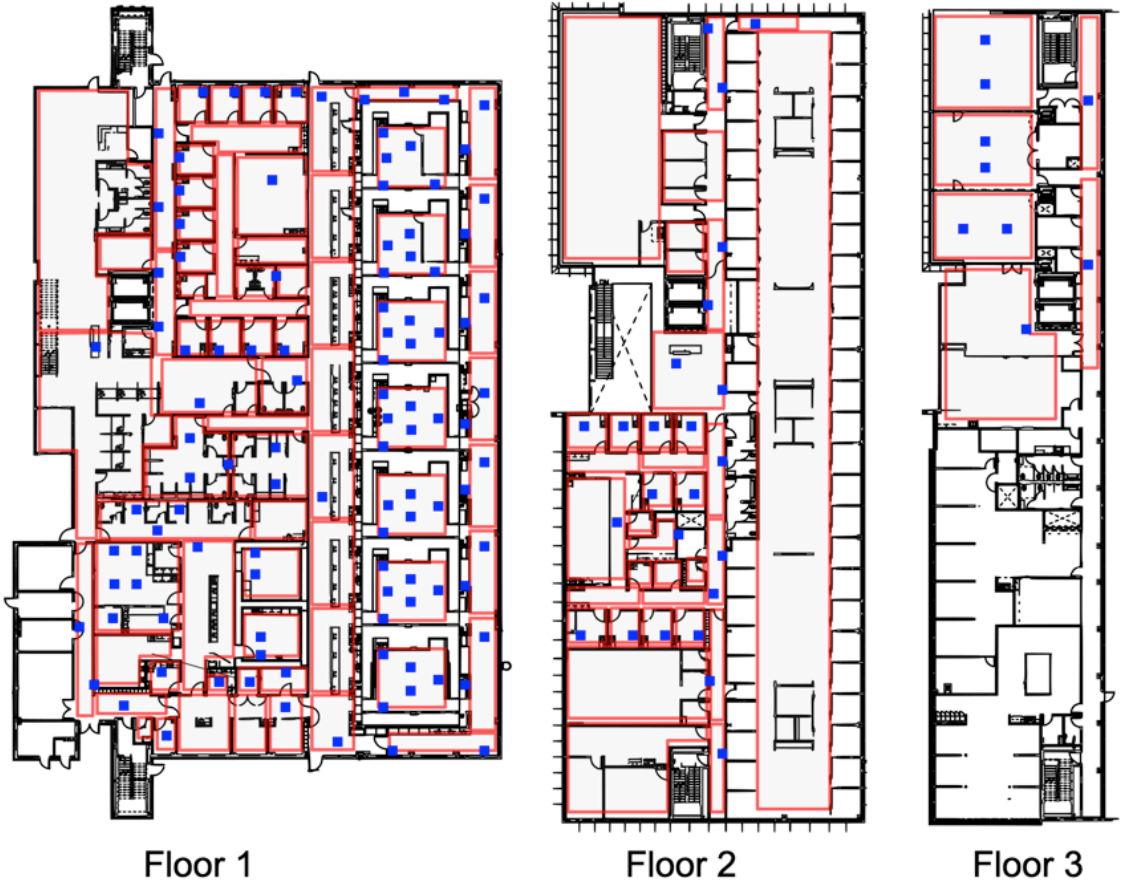

**Figure 2.** Floor plan of EROC with 115 zones (red-board boxes) and 142 BLE scanners (blue squares). A zone is typically an exam rooms, a treatment room, an open office area, or a waiting area. The scanners are installed in the ceiling so that are invisible from the ground.

Figure 3 illustrates the detailed structure of the proposed network, which includes an embedded LSTM unit and a deep classifier. The LSTM unit used in our method is a multi-layer unit taking signal strength measured from multiple sensors as input and outputs two vectors as historical information passed to the next time step of which one is fed into the deep classifier. More precisely, the LSTM unit can be explicitly expressed as the following equations [27]:

$$i_t = \sigma(W_{xi}x_t + W_{hi}h_{t-1} + W_{ci}C_{t-1} + b_i), \qquad (1.1)$$

$$f_t = \sigma(W_{xf}x_t + W_{hf}h_{t-1} + W_{cf}C_{t-1} + b_f), \qquad (1.2)$$

$$C_t = f_t C_{t-1} + i_t \tanh(W_{xc} x_i + W_{hc} h_{t-1} + b_c), \tag{1.3}$$

$$o_t = \sigma(W_{xo} x_t + W_{ho} h_{t-1} + W_{co} C_t + b_o), \tag{1.4}$$

$$h_t = o_t \tanh(C_t). \tag{1.5}$$

$W_{xi}, W_{xf}, W_{xc}, W_{xo} \in \mathbb{R}^{200 \times 142}$, $W_{hi}, W_{hf}, W_{hc}, W_{ho}, W_{ci}, W_{cf}, W_{co} \in \mathbb{R}^{200 \times 200}$, and $b_i, b_f, b_c, b_o \in \mathbb{R}^{200}$ are weight matrices and bias vector parameters of fully connected layers to be learned from the training. Subscript $t$ denotes the time step. $C_t, h_t, i_t, o_t, f_t \in \mathbb{R}^{200}$. $x_t \in \mathbb{R}^{142}$ is the input RSSI (Received Signal Strength Indicator) vector indicating the measured signal strength. $C_t$ denotes the state of the network cell containing the memory to be passed to the next time step, and $h_t$ denotes the output vector which is fed into the deep classifier and also passed to the consecutive time step. $f_t, i_t, o_t$ are activation vectors for the forget gate, input/output gate, and output gate, respectively, which control the extent to which the cell and information passed from previous steps are used in the calculation. $\sigma$ is the sigmoid activation function. The purpose of the LSTM unit is to take into account the historical information in determination of the current location and to pass the current location information to later steps.

The deep classifier is composed of 2 fully-connected dense layers, each with a rectifier linear unit (ReLu) activation [28] and a softmax layer [29] to calculate the probability of each zone. Each fully-connected layer was followed by a dropout layer [30] with a fraction of 0.5 in order to reduce overfitting.

The output of the deep classifier is a vector of which each element represents the probability of a zone. Finally, the zone with the greatest probability is taken as the result of location determination.

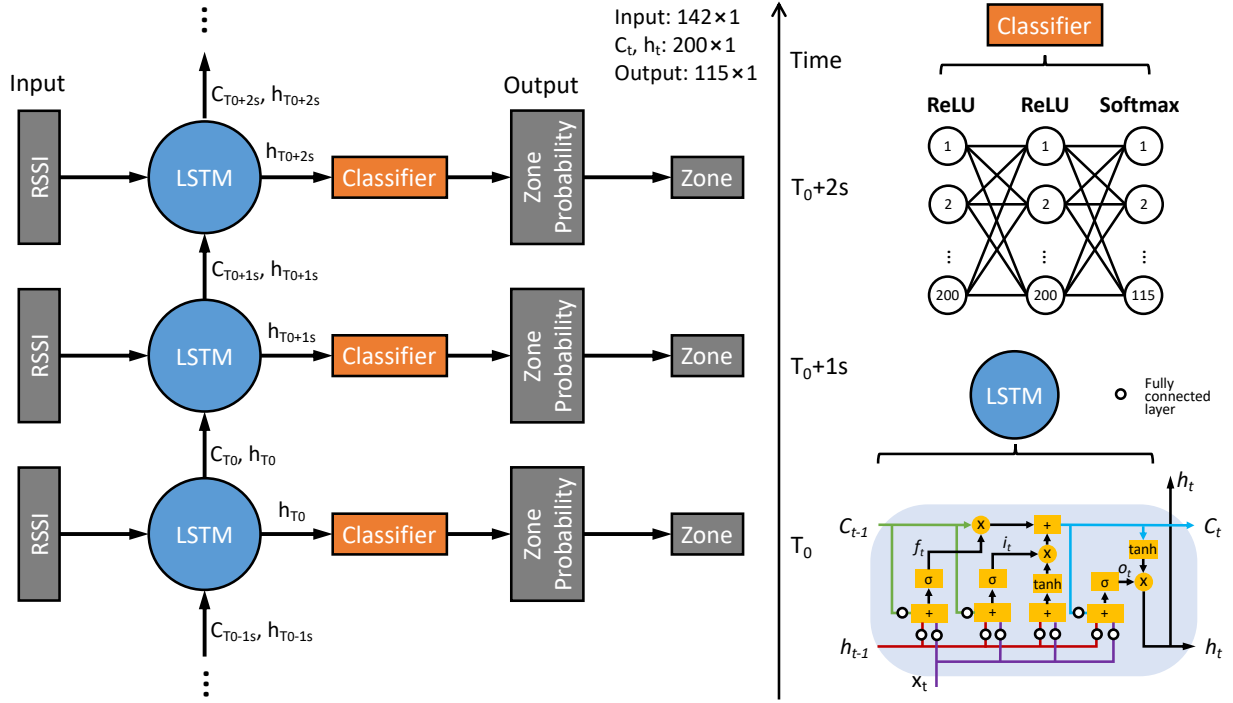

**Figure 3**. Illustration of the LSTM-classifier Neural Network for location computation. Input is a sequence of size 142 vector where each element represents the RSSI from a sensor. The LSTM unit outputs two size 200 vectors, one passed to the classifier, and both passed to the next time step. Classifier consists of two fully-connected dense layers each having 200 neural cells. Output after Classifier layer is a vector of size 115 corresponding to the probabilities of 115 zones. Final output is the zone number with the maximum probability.

The training of the network is the process to determine all network parameters via minimizing the following loss function on training dataset:

$$L(\mathbf{w}) = -\frac{1}{N}\sum_{i=1}^{N}[y_i \log(\hat{y}_i) + (1-y_i)\log(1-\hat{y}_i)], \qquad (2)$$

where $\mathbf{w}$ denotes the model parameters, $N$ is the number of samples, $y_i$ is the true zone number, and $\hat{y}_i$ is the predicted zone number. Stochastic gradient descent algorithm was implemented by solving optimization problem:

$$\mathbf{w}^n = \mathbf{w}^{n-1} - \alpha \frac{\partial L}{\partial \mathbf{w}}, \qquad (3)$$

where superscript $n$ is the iteration number.

*2.4 Posterior constraint*

The network proposed in last subsection could become ambiguous in trajectory tracking during zone transitions when the tracked object is closed to the boundary of zones. In such cases, as mentioned in the Introduction, when the latency is considered less of a concern, we used the spatial relation between zones to regularize the object from moving to a far-away zone in a very short period of time, so as to increase stability. Therefore, we proposed to postprocess the network's output (a vector of zone probabilities) to use the previous zone and zone connection information to constrain the current zone:

$$P(Z_t^n | Z_{t-\Delta t}^m) = \frac{P(Z_t^n) P(Z_{t-\Delta t}^m | Z_t^n)}{P(Z_{t-\Delta t}^m)}, \tag{4}$$

where $P(Z_t^n)$ is the probability of the $n^{th}$ ($n$=1,2,…,115) zone calculated from the model at time $t$, $P(Z_{t-\Delta t}^m)$ is the probability of the $m^{th}$ ($m$=1,2,…,115) zone in the previous time step $t - \Delta t$, $P(Z_t^n | Z_{t-\Delta t}^m)$ represents the probability of the $n^{th}$ zone at time $t$ under the condition that the previous zone is $m$, and $\Delta t$ is the time difference between the current timestep and the previous timestep in seconds. For simplicity, we assumed the predicted zone of the previous step is known (assuming taking the zone with the maximum probability), so that

$$P(Z_{t-\Delta t}^m) = 1. \tag{5}$$

The conditional probability $P(Z_{t-\Delta t}^m | Z_t^n)$ is assumed as

$$P(Z_{t-\Delta t}^m | Z_t^n) = \gamma \cdot k^{-d_{mn}/\Delta t}, \qquad (6)$$

where $d_{mn}$ is the "distance" between the $n^{\text{th}}$ zone (candidate current zone) and the $m^{\text{th}}$ zone (previous zone) and is defined as the minimum number of crosses between two zones, e.g., distance between two connected zones is 1, and between the same zone is 0 when $m=n$, $k$ is a parameter inferring how fast the tracked object can move, and $\gamma$ is the normalization factor such that $\sum_n P(Z_t^n | Z_{t-\Delta t}^m) = 1$. $P(Z_t^n)$ is directly calculated from the deep classifier. Eventually, the maximum of $P(Z_t^n | Z_{t-\Delta t}^m)$ is taken to predict the zone at time *t*. Equation (6) indicates that it is less probable for a patient to move to a zone far away from the previous zone, but this spatial-probability-constraining effect diminishes over time. For example, the longer the time since the last known location, the more likely that the patient will show up in a zone far from the previous known zone. Note that errors in computing the previous zone can affect the current zone computation, because, in real-time prediction, the ground truth is unknown for both the current and the previous steps. Thus, parameter *k* should be carefully optimized so that the constraint is effective and also that the effect of errors from the previous location calculation can be reduced when calculating the current zone. In particular, *k*=1 infers that the object can move very fast so that historical location won't constrain the current determination at all, and *k*=∞ infers moving very slowly so that it always stays in a zone. Additionally, the minimum number of crossings between two zones ($d_{mn}$) is not exact but an approximation of their distance. However, we used such a posterior constraint scheme that relies heavily on approximation along with the LSTM unit and deep classifier to predict the trajectory, and we reasonably expected to see an improvement in stability. Figure 4 illustrates the proposed scheme composed of an LSTM unit, a deep classifier, and a posterior constraint scheme.

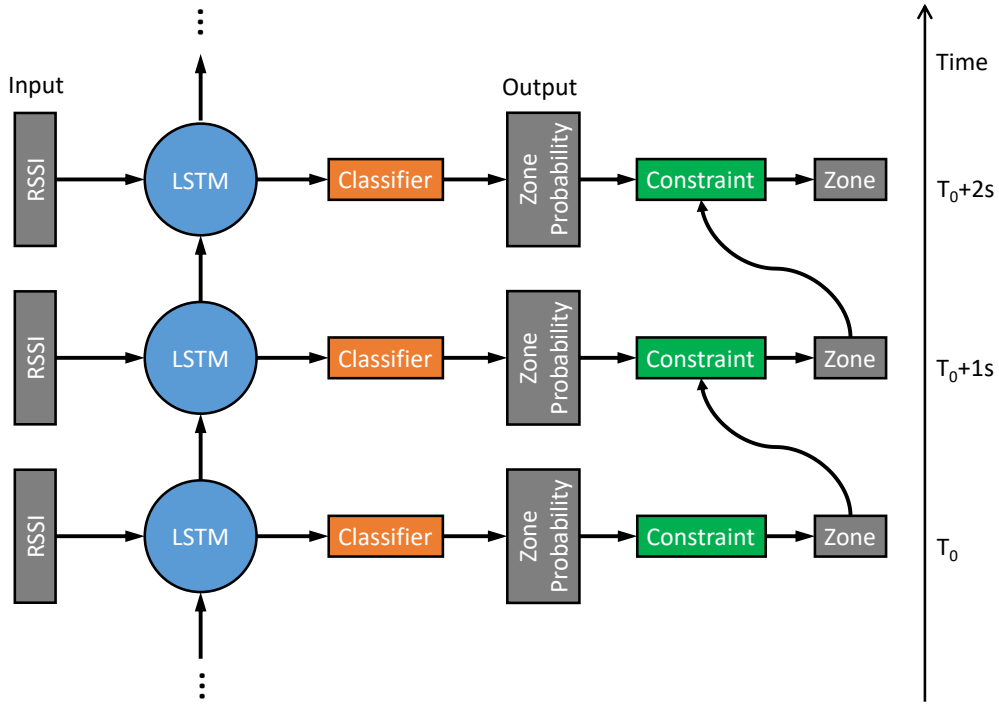

**Figure 4**. Illustration of the proposed algorithm for location computation. Layers Input, LSTM, Classifier, and Output are the same as in Fig. 3. Constraint layer uses calculated zone from the previous step to calculate the conditional zone probability before taking the maximum (following Eqs. 4-6). Final output is the zone number with the maximum conditional probability.

*2.5 Data collection and experiments*

2.5.1 Data collection

To collect training and testing data, we carried beacon tags around each zone and collected the raw signal strength data and corresponding zone numbers as labels. The tags were carried in front of the chest using a lanyard, in the shirt pocket, or in the pant pocket to mimic different scenarios for real patients. We walked randomly but tried to go through all positions and corners in each zone to reduce gaps between classes. In the first experiment, we carried 6 tags in different body positions to collect the training set and carried 1 additional tag to independently collect the testing set in a different trip. For both training and testing, for each zone and each tag, we collected data for 2 minutes with approximately 1200 timesteps considering the BLE broadcast interval of 0.1 second. In the second experiment, we collected 10 trajectories during which we carried the tag across zones in order to valid and test trajectory tracking. Each trajectory was about 20 minutes and encountered 10-20 cross-zone events. The ground truth locations were labeled using real-time location.

2.5.2 Data preparation

Trajectory tracking requires collecting training data that includes cross-zone trajectories. However, this is beyond our capability because there could be too many trajectories to collect, and the labeling process is also tedious. To compensate, we collected labeled data in each zone separately and augmented them to simulate training trajectories for the proposed network. Each simulated trajectory consisted of $2N$ seconds data of which the first $N$ seconds were randomly selected from one zone and the last $N$ seconds selected from one of its connected zones. Note that the selected $N$ seconds data were consecutive in each zone. We thus did such augmentation for every pair of connect zones. Though not perfect, we considered the simulated trajectories as relative realistic ones because they are only between connected zones, and we used them to train the proposed algorithm to track moving objects.

2.5.3 Experimental setup

We deployed a fleet of 142 BLE sensors to locate BLE tags in a 3-floor building which is divided into 115 clinically meaningful zones. The BLE tags broadcast signals every 0.1 second, which is received by the sensors and transferred to a server to calculate the location. To stabilize the signal, the signal strengths measured from the sensors were averaged over the past $\Delta t = 1$ second, and the interval between two consecutive steps of input trajectories to the proposed algorithm was also $\Delta t = 1$ second.

To train the algorithm to track cross-zone trajectories, we augmented the data to simulate a total of 3870 trajectories. For each of the 6 training tags, for each of the 215 pairs of connected zones, we simulated 3 trajectories each consisted of 50 seconds of which the first $N=25$ seconds were randomly selected from one zone and the last $N=25$ seconds were from the other zone. These generated sequences were then fed into the proposed algorithm to tune hyperparameters.

We tuned the parameter $k$ for the posterior constraint scheme and the so-called lookback parameter, which indicates the length of history that LSTM looks back, by validating over randomly selected half of the 10 collected trajectories. The rest half trajectories were then used as the testing set. In our settings, the size of lookback is the same as history length in seconds.

Among the 6 tags for training, we randomly selected 5 as training set and the rest 1 as validation set. To train the proposed network, we fed the training set to it and iterated following Eq. 3 over a large epoch number. The validation loss function was output during each iteration. The parameters were then selected at the epoch number when the validation loss function reaches minimum.

We trained the proposed network on a CPU using TensorFlow 1.13.1 and Keras (CPU version) 2.2.4 of Python 3.5 hosted by a Dell Alienware desktop with an i7-8700 CPU. We trained CNN+ANN model on the same CPU and software following the procedures suggested by [12], except that the input dimension was switched from two to three and the output size was switched from 21 to 114 to adapt to the multi-floor building with more zones.

*2.6 Evaluation*

For most applications, a delay of a few seconds is less of a concern compared to the accuracy and stability, thus, it is meaningful to define the accuracy of trajectory tracking in a way that the effect of delay if neglected. Therefore, we defined that the determined location at a certain time step of a trajectory is correct as long as it is the same as any of the ground truth locations within preceding and following 10 seconds, which is an empirical number of seconds. Such a definition eliminated the effect of delay in the accuracy evaluation. However, if the delay is too much, e.g., over 10 seconds, we still counted it as incorrect prediction.

Besides of the defined accuracy, we used the number of incorrect zone changes to quantify the stability of the algorithm. We defined a zone change to be correct if the same change occurs in the ground truth trajectory within preceding and following 10 seconds, and to be incorrect if the same change didn't occur in the ground truth trajectory within preceding and following 10 seconds, or the correct change has already taken place. The algorithm is considered more stable if it results in smaller number of incorrect zone changes for trajectory tracking.

## 3. Results

We first evaluated the proposed network in terms of accuracy for individual zones, neglecting cross-zone events. We tuned the network parameters by training and validating using the training and validation sets, respectively. Figure 5 shows the training and validation loss functions of the proposed network as a function epoch. The training loss function continually decreased over epoch while the validation loss function decreased and then saturated after about 5-10 epochs. We selected the parameters at 23rd epoch when the validation loss function is the smallest. For CNN, we observed similar trends of the training and validation loss functions and selected the parameters at the 32nd epoch when the validation loss function is the smallest. For ANN, the loss validation function took 2 epochs to saturate, and we selected parameters at the 2nd epoch. The selected parameters were then used to calculate the location of the testing set.

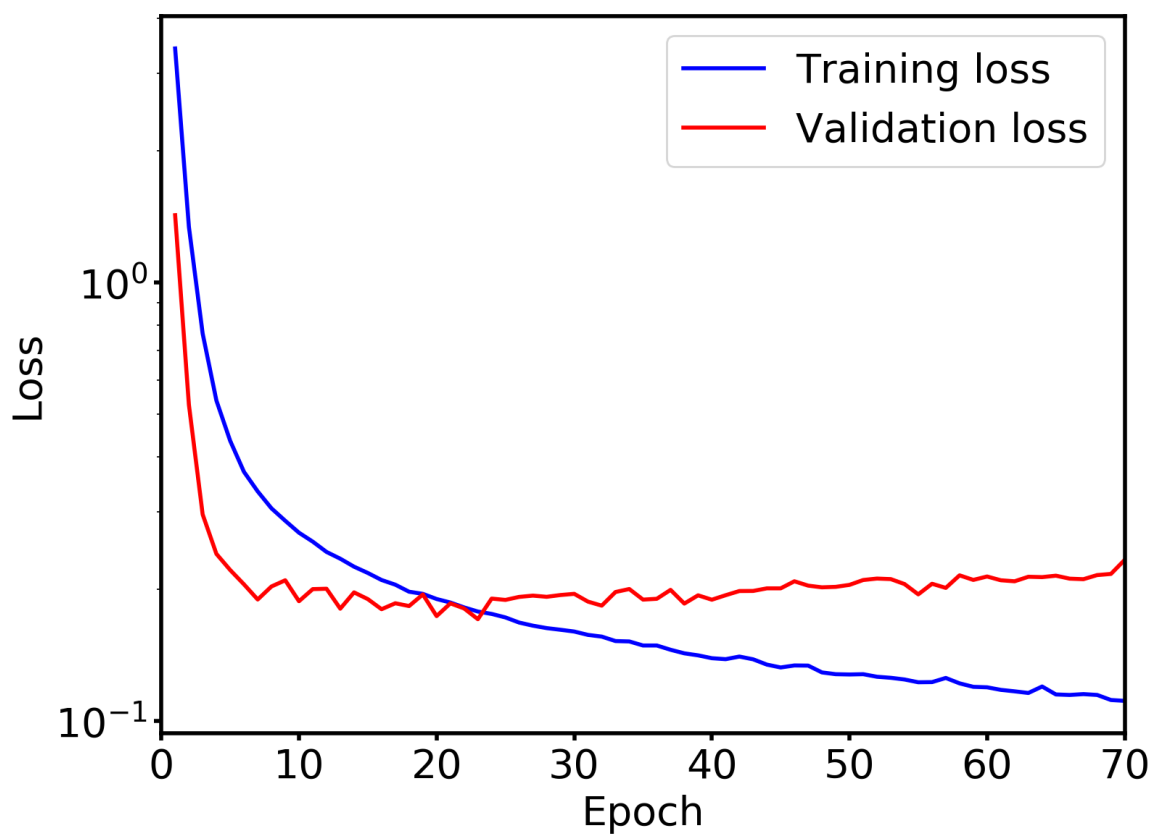

**Figure 5.** Loss function of the training set and validation set for the proposed network.

Figure 6 shows the location accuracy of testing data for the proposed network and CNN+ANN as functions of the history length. The proposed network outperforms CNN+ANN at the same history length. Accuracy of the proposed network saturates when the history length is about 10. Note that the CNN+ANN scheme in our experiment is less accurate than in the initial report [12], presumably because that paper focused on a much smaller area where the sensors are relatively dense; the zones in that paper were also larger than in our experiments.

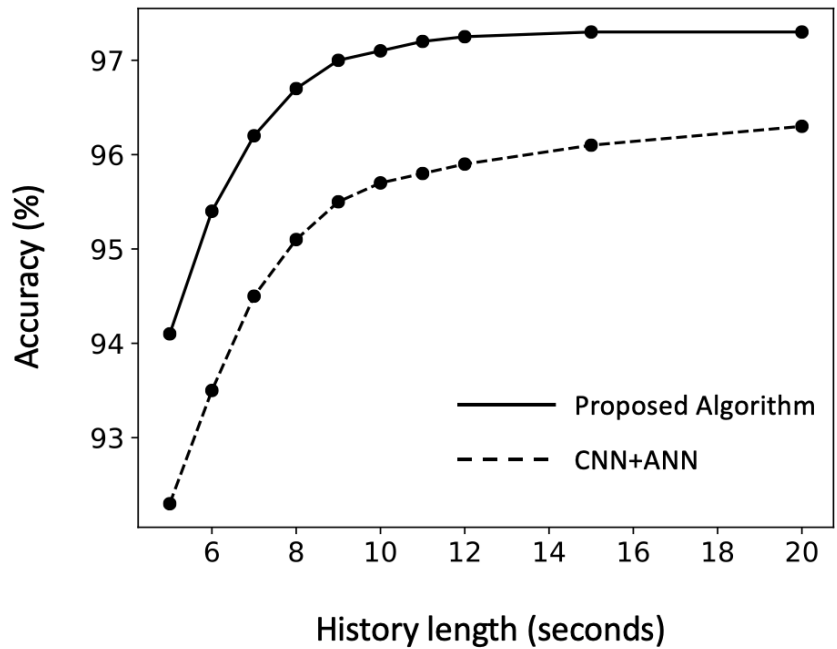

**Figure 6**. Location accuracy of the proposed network and CNN+ANN as functions of history length. Training and testing samples included no cross-zone trajectories.

To give a sense how many sensors are necessary for accurate localization, we quantified the relationship between sensor density and accuracy. We defined the sensor density as the total number of sensors divided by the total number of zones. Here we did not use the intrinsic definition of density (number of sensors divided by total floor area) considering that larger zones are more easily localized than smaller zones, but it is not reflected in such a definition. Thus, we considered #sensor / #zone rate as a more reasonable metric for sensor density. We randomly removed certain number of sensors before training the proposed network to obtain a relationship between accuracy and sensor density, which is shown in Fig. 7. The figure shows that the accuracy saturated when the sensor density reached about 1.0, meaning that to fully explore the BLE technology in indoor localization, the number of sensors should be close to the number of zones. The accuracy reduces to about 90% when the number of sensors is half of the number of zones. Further reducing sensor density will lead to accelerating inaccuracy.

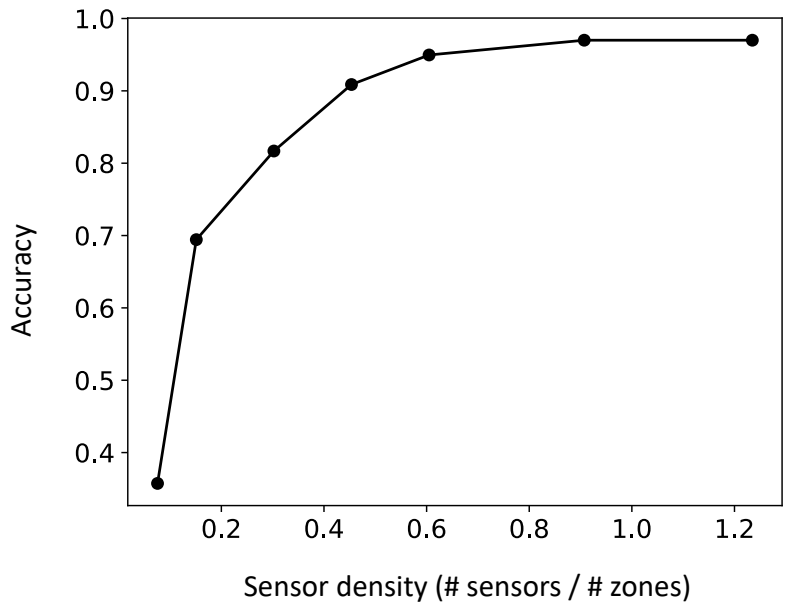

**Figure 7**. Accuracy of the proposed network as a function of sensor density. The sensor density is defined as total number of sensors divided by total number of zones. A history length of 10 seconds was used.

For the inaccurate classification results, the question naturally emerges how far the classification results are from the ground truth. An inaccurate result may still be acceptable for certain applications if it is close to the ground truth, that is, if it is "not so wrong." To quantify this, we calculated the distance between the classification results and the ground truth following the definition of $d_{mn}$ in Equation (6). The average distance is 1.27. Note that a distance of 1 means that the two zones are directly connected and close in space, and some spatially-close zones are unconnected (e.g., two rooms separated by a wall) and thus have a larger distance. Therefore, we conclude that the wrong classification results are still relatively spatially close to the ground truth.

They are "not very wrong." In addition, because most wrong classifications happened in clinically unimportant areas, they wouldn't significantly affect clinical applications.

We used the accuracy and stability defined in Section 2.6 to evaluate the trajectory tracking of the proposed algorithm. We first tuned the network parameters, the parameter $k$ and the history length using the training set, which were simulated using the collected data from individual zones following the procedures in Section 2.5, and the validation set, which were 5 trajectories randomly selected from the 10 collected trajectories. Table 1 shows the accuracy and average number of incorrect zone changes for the proposed algorithm for the validation cross-zone trajectories. Various parameter $k$ values and lookback sizes were examined. Note that $k=1$ corresponds to no posterior constraint, and smaller number of incorrect zone changes indicates better stability. Both accuracy and stability increased with k and lookback size. The proposed algorithm with $k=40$ and lookback size of 10 has the best stability and accuracy. Note that the accuracy here was defined so that the effect of delay was eliminated.

Table 1 Accuracy and average # incorrect zone changes of the proposed algorithm for various $k$ value for the validation cross-zone trajectories. Accuracy* was defined in Section 2.6 where the effect of latency was eliminated.

| K | Lookback = 10 | | Lookback = 6 | |
|---|---|---|---|---|
| | Accuracy* | Average # incorrect zone changes | Accuracy* | Average # incorrect zone changes |
| 1 | 96.1% | 11 | 95.4% | 15 |
| 5 | 97.2% | 6 | 96.9% | 10 |
| 10 | 98.8 | 2 | 97.5% | 5 |
| 40 | 100% | 0 | 98.6 | 3 |

To test the generality, we selected parameter *k* to be 40 and lookback size to be 10 for proposed algorithm and tested on the testing dataset. The testing results were shown in Table 2 along with the CNN+ANN model results. The proposed algorithm still achieved an accuracy of 100%, outperforming CNN+ANN. As an example, Fig. 8 shows a testing trajectory collected in the exam area in the first floor predicted by both the proposed algorithm and CNN+ANN. The latter resulted in 10 wrong cross-zone events. Most of the wrong zone changes were due to the ambiguity of the algorithm at the zone boundaries.

Table 1 Accuracy and average # incorrect zone changes of the proposed algorithm and CNN+ANN for the testing cross-zone trajectories. The history length was 10 seconds for both methods. Accuracy* was defined in Section 2.6 where the effect of latency was eliminated.

| Methods | Accuracy* | Average # incorrect zone changes |
| --- | --- | --- |
| CNN+ANN | 94.5% | 12 |
| Proposed algorithm | 100% | 0 |

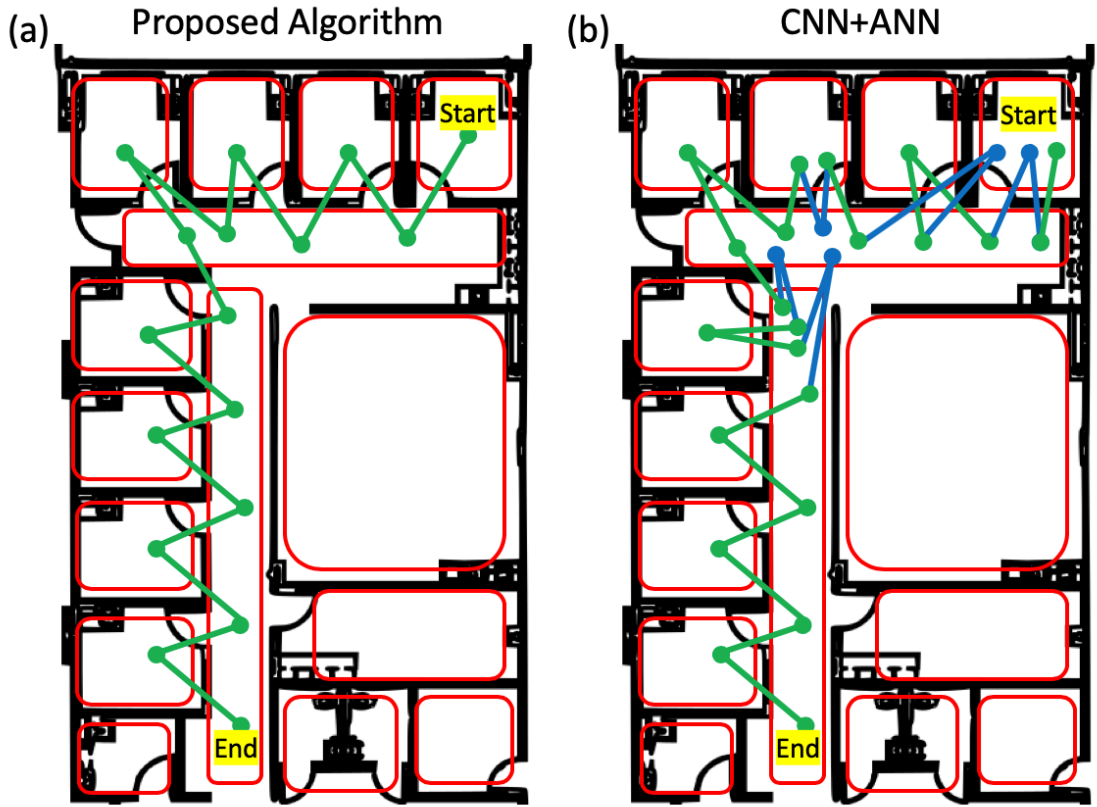

**Figure 8**. Predicted results for a testing trajectory collected on the first-floor exam area starting from the top-right exam room and ending at the bottom-left corridor by (a) the proposed algorithm and (b) CNN+ANN. Green lines represent correct cross-zone events, while blue lines represent wrong cross-zone events, and red squares represent zones.

## 4. Discussion and Conclusions

We proposed a deep neural network to increase the accuracy and stability for a BLE-based indoor Real-time Location System (RTLS) and evaluated it in a testbed built in the Department of Radiation Oncology at the University of Texas Southwestern Medical Center. The proposed algorithm, composed of an LSTM architecture, a deep classifier, and a posterior spatial-temporal constraint scheme, has substantially increased the stability and accuracy of RTLS.

To train the network to track moving objects, we augmented the training data collected in individual zones to simulate cross-zone trajectories using the spatial connection information of zones. To further increase stability, we explored the spatial relation between zones and postprocessed the location results to consider the previous location as a constraint condition. These efforts together resulted in better stability and accuracy when the latency is less of a concern.

From our experiments, the total number of sensors should be similar to or greater than the total number of zones to fully explore the strength of the deep learning technique. An easy implementation strategy would be to install one sensor in each zone.

We have made many assumptions and approximations in our schemes. The distance between two zones defined in this paper is not precise though briefly reflects the real distance. The previous zone was assumed to be known using the predicted value in the previous time step in the posterior constraint scheme. Considering this shortcoming, a Kalman filter may be better at stabilizing the location results than the posterior constraint scheme.

Infrastructure and environments change over time, and this will affect the function mapping the signal footprint to location. For example, the EROC building has built and demolished a few inner walls since we collected the training data. This affects the real-time testing accuracy, especially for zones close to the changed walls. Thus, we consider periodically re-collecting training data and re-training the model.

The data and code used in this paper were shared at https://github.com/S184490/RTLS.

## Acknowledgement

The authors acknowledge Erlei Zhang for useful discussion of algorithms. The authors thank Jonathan Feinberg for editing the manuscript.